\documentclass{pasj00}

\begin{document}
\SetRunningHead{K. Watarai and S. Mineshige}
{Marginally Stable Orbit in Super-Critical Accretion Flow}
\Received{2003/06/03}
\Accepted{2003/08/20}

\title{Where is a Marginally Stable Last Circular Orbit in Super-Critical
Accretion Flow?}

\author{Ken-ya \textsc{Watarai}\altaffilmark{} %
and Shin \textsc{Mineshige}\altaffilmark{} }

\email{watarai@yukawa.kyoto-u.ac.jp}

\altaffiltext{}{ Yukawa Institute for Theoretical Physics, Kyoto University,
        Sakyo-ku, Kyoto 606-8502}



\maketitle

\begin{abstract}
Impressed by the widespread misunderstanding of the issue, we return to
the old question of the location of the inner edge of accretion disk
around black hole. We recall the fundamental results obtained in the
1970's and 1980's by Warsaw and Kyoto research groups that proved, in
particular, that the inner edge does not coincide with the location of the
innermost stable Keplerian circular orbit. 
We give some novel illustrations of this particular point and of some
other fundamental results obtained by Warsaw and Kyoto groups.
To investigate the flow dynamics of the inner edge of accretion disk,
 we carefully solve the structure of the transonic flow and plot
the effective potential profile based on the angular-momentum
distribution calculated numerically. 
We show that the flow does not have a potential minimum
for accretion rates, ${\dot M}\gtrsim 10 L_{\rm E}/c^2$  
(with $L_{\rm E}$ being the Eddington luminosity and 
 $c$ being the speed of light). 
This property is realized even in relatively small viscosity parameters 
 (i.e., $\alpha \sim 0.01$), because of the effect of pressure gradient. 
In conclusion, the argument based on the last circular orbit of a test
 particle cannot give a correct inner boundary of the super-critical
 flow and the inner edge should be determined in connection with
 radiation efficiency. 
The same argument can apply to optically thin ADAF.
The interpretation of the observed QPO frequencies should be
 re-considered, since the assumption of Kepler rotation velocity can
 grossly over- or underestimate the disk rotation velocity, depending
 on the magnitude of viscosity. 
\end{abstract}

\section{Introduction}
For the reason already explained in the Abstract, we return to the old
question where the inner edge $r_{\rm in}$ of black hole accretion disk is
located. This, and other questions concerning the transition region where
the flow of accreted matter changes its character from being almost
circular to being almost free-fall, have been intensely studied, and most
of them definitely answered, in the late seventies and early eighties by
two research groups: the Warsaw group lead by Bohdan Paczy{\'n}ski, and by
the Kyoto group lead by Shoji Kato.

Researchers in Warsaw and Kyoto formulated several exact, simple, and
practically useful analytic theorems and formulae, that were very much
independent of the nature of dissipation, radiative processes and
viscosity. They gave a general description of the innermost part of black
hole accretion flow, i.e. the region located between the black hole
horizon $r_{\rm g}$ and the marginally stable circular Keplerian orbit
$r_{\rm ms}$.
Later developments in hydrodynamics and magnetohydrodynamics of accretion
disks have brilliantly confirmed these results. This includes detailed
numerical models of supercritical (accretion rate higher than the
Eddington one), optically thick and geometrically either thick or slim
accretion disks, and sub-critical, optically thin and geometrically thick
ADAFs and iron tori (see Kato et al. 1998 for a review of these models).        

Questions concerning the transition region and the inner edge of the disk
cannot be properly addressed in the ``standard'' Shakura -- Sunyaev thin
accretion disk model because of two crucial mathematical simplifications
that the standard model makes --- that the angular momentum of the
accreted matter $\ell(r)$ is everywhere Keplerian, $\ell(r) =
\ell_{\rm K}(r)$, and
that the viscous heating is locally balanced by radiative cooling. In
outer parts of the disk, $r \gg r_{\rm ms}$, these assumptions are quite
acceptable in the sense that several calculated disk's properties, in
particular the spectra, do not depend much on them. They fail completely
in the innermost part of the flow, $r < r_{\rm ms}$, where they give wrong
qualitative picture of the flow.

Abramowicz and Kato (1989) gave a useful, short and adequate summary of
the early Warsaw and Kyoto research with explicit quotations to authors of
particular results. Here in this Introduction we repeat three points of
their summary in a different order and wording, and without references.
All the statements below are general, proven, theorems confirmed later by
3D numerical supercomputer simulations (see Igumeshchev et al. 2003 for
references to papers describing results of these simulations):

[1] Dynamics. In the innermost part of the disk angular momentum of
accreted matter is not Keplerian. The dynamics of the flow depends mostly
on the magnitude of the angular momentum there: flows with low angular
momenta have very different properties from those with high angular
momenta.

The low angular momentum, Bondi-type flows have $\ell(r) < \ell_{\rm ms} =
\ell_{\rm K}(r_{\rm ms})$. The centrifugal force is dynamically unimportant
everywhere, and the flow is far from mechanical equilibrium. Flow lines
are never close to circles, and the flow properties are similar to these
known from the case of a spherical accretion. For the Bondi-type flows,
the very concept of the inner edge loses its usefulness: in the
Bondi-type of accretion flows there is simply no characteristic radius
that could be called the inner edge of the flow.

The high angular momentum, disk-type accretion flows have, $\ell(r) <
\ell_{\rm ms}$. The outer part of the flow, i.e. the accretion disk itself, is
close to mechanical equilibrium determined by gravitational, centrifugal,
and pressure forces. The total gravitational plus centrifugal potential
$\psi_{\rm eff}(r, z)$, has two extrema at locations $r_*$ given as the two
solutions of the equation $\ell(r_*) = \ell_{\rm K}(r_*)$,
 i.e. where the angular
momentum of the matter $\ell(r_*)$ equals to the corresponding Keplerian
value at the same radius $r_*$. Note, that because the relativistic
Keplerian angular momentum distribution has a minimum at $r_{\rm ms}$
 and we are now interested in the case $\ell(r) > \ell_{\rm ms}$,
 there must be two such locations on both sides of $r_{\rm ms}$. 
The first one, $r_* = r_{\rm cent} > r_{\rm ms}$,
 corresponds to a minimum of $\psi_{\rm eff}(r, z)$ and a maximum of the
pressure. 
The second one, $r_* = r_{\rm cusp} < r_{\rm ms}$ corresponds to a
maximum of $\psi_{\rm eff}(r, 0)$ at the equatorial plane $z = 0$. 
The equipotential surface
 $\psi_{\rm eff}(r, z) = {\rm const} = \psi_{\rm eff}(r_{\rm cusp}, 0)$
crosses itself along the circle $(r, z) = (r_{\rm cusp}, 0)$. 
This self-crossing equipotential is called the Roche lobe. 
Inside the disk, i.e. for $r > r_{\rm cusp}$,
 matter very slowly moves inward due to the action of viscous
torques. 
For $r < r_{\rm cusp}$ it goes inward very fast, indeed
transsonically, due to the insufficient centrifugal support and therefore
lack of mechanical equilibrium. 
This is similar to the more familiar situation of the Roche lobe
overflow known in the context of close binaries. 

Thus, in the region close to the location of the cusp,
 $r = r_{\rm cusp}$, the flow changes its character from almost
 circular to almost free-fall, 
 and for this reason, the cusp represents exactly what it is the crucial
physical meaning of the ``inner edge'' of the disk. The other possibility
would be the sonic radius $r_{\rm sonic}$, where the inward radial velocity of
the flow changes from sub- to super- sonic. 
The sonic radius is just slightly closer to the black hole than the
 cusp, $r_{\rm mb} < r_{\rm sonic} < r_{\rm cusp} < r_{\rm ms}$.  

[2] Viscosity and two types of the flow. For stationary accretion flow
that asymptotically, i.e. for large radii, are described by the standard
Shakura -- Sunyaev model, and have viscosity given by the standard
$\alpha$ prescription, the value of the angular momentum in the inner part
of the disk, expressed in the dimensionless units of $\ell_K(r_{\rm ms})$,
depends on $\alpha$ and accretion rate ${\dot M}$ 
(Muchotrzeb and Paczy{\'n}ski 1982; Matsumoto et al. 1984).

For any value of ${\dot M}$ there is a critical value of $\alpha_{\rm crit}$
such that for $\alpha > \alpha_{\rm crit}$ the flow is of the Bondi type,
while for $\alpha > \alpha_{\rm crit}$ the flow is of the disk type. For small
very subcritical accretion rates $\alpha_{\rm crit} < 0.1$

[3] Location of the inner edge and efficiency. Assuming ``no torque inner
boundary conditions'', i.e. that viscous torque cannot act across the
black hole horizon, location of the cusp is directly connected to the
efficiency of accretion. The efficiency equals to minus the binding
orbital energy at the inner edge divided by speed of light square, $\eta =
- e_{\rm K}(r_{\rm in})/c^2$. In particular, for accretion disks that are
radiatively inefficient $\eta \approx 0$,
 i.e. thick disks, slim disks and ADAFs,
 the inner edge approaches the location of the marginally bound circular
 Keplerian orbit $r_{\rm mb}$, because $\eta(r_{\rm mb}) = 0$. 
For the Schwarzschild i.e. non-rotating black hole with the mass $M$,
 it is $r_{\rm g} = 2GM/c^2$, $r_{\rm mb} = 2 r_{\rm g}$, and $r_{\rm
ms} = 3 r_{\rm g}$. 

These fundamental results of Warsaw and Kyoto groups have been reviewed
many times, also in the best known textbooks and monographs on accretion
theory (Frank, King and Raine 2002 3rd edition; Kato, Fukue and Mineshige
1998; Abramowicz, Bj{\"o}rnsson and Pringle 1998). Some of them belong to
the most often quoted results in the black hole accretion disk theory. It
is therefore quite surprising, that still today there is a widespread and
wrong conviction that the innermost edge of the accretion disk must
coincide with the location of the marginally stable circular Keplerian
orbit $r_{\rm ms}$. 
This wrong opinion appears to be especially popular among observers. 
While it is quite possible (as often suggested) that
the highest frequency of quasi periodic variability observed in many of
these sources may come from blobs of matter orbiting the central black
hole or neutron star close to the inner edge of the disk, one cannot claim
that the observed highest frequency equals the orbital frequency at the
marginally stable circular Keplerian orbit.

This article is another attempt (see Paczy{\'n}ski, 1998) to clear the
issue of the inner edge of the disk by reminding the classic early results
of Warsaw and Kyoto groups. Here we are discussing some new illustrations
of them. We do not always quote names of the original Warsaw and Kyoto
authors of particular classical results. 

In this paper, we thus discuss the stability of circular orbits. 
In section 2, we explicitly show the shape of the effective potential
and examined the stability.  
In section 3, we discuss the unique properties of the transonic
regions of the slim disk and their observational implications. 
The last section is devoted to conclusions.

\section{Effective Potentials of Accretion Disk}

\subsection{Basic Considerations}
It has been shown by the X-ray observations of
 black hole candidates (BHCs) that the radius of the inner boundary
estimated through the spectral fitting to the X-ray data is constant in time 
despite large-amplitude variations in luminosity, and 
it coincides with $3~r_{\rm g}$ within error bars
 (e.g. Makishima et al. 1986; Tanaka \& Shibazaki 1996; Ebisawa 1999).

Caution should be taken regarding this argument, however, since
all the accreting material should eventually pass through the
unstable regions before finally falling across the event horizon.
That is, certain amount of materials should at any times exist 
in the unstable regions.
(This contrasts the case of a static system, in which we may argue
that there are practically no materials present in unstable regions.) 
Although matter density is non-zero inside $3~r_{\rm g}$, 
the unstable region has no observational significance for BHCs 
with luminosity being well below the Eddington luminosity,  
since these regions are optically thin, thus emitting fewer photons. 
Some observational effect appears when the luminosity becomes
comparable to the Eddington luminosity, for which the flow inside
$3~r_{\rm g}$ might become optically thick, thus producing significant
soft X-ray emission (Watarai et al. 2000). 
Some ULXs (ultra-luminous X-ray sources) show a trend that 
the apparent inner-edge of the disk decreases with increase of 
luminosity (Mizuno, Kubota, \& Makishima 2001;
 Watarai, Mizuno, \& Mineshige 2001). 
Microquasar GRS1915+105 exhibits rapid changes of the inner edge
radius during transitions (Belloni et al. 1997; Yamaoka et al. 2001
; Watarai \& Mineshige 2003). 

It has been claimed that usual arguments regarding the inner boundary 
of the disk based on the particle orbits are sometime misleading in the 
transonic flow (see Kato, Fukue, \& Mineshige 1998 Chapter 2 for a review).  
To see this problem more clearly, let us first recall the basic
argument on 
the stability of the circular orbits around a non-rotating black hole. 
General relativistic description of the effective potential leads 
\begin{equation} 
\psi_{\rm eff}^{GR}(r) = 
  -\frac{GM}{r}\left(1+\frac{\ell^2}{c^2 r^2}\right) + \frac{\ell^2}{2r^2}
\label{eq:p0}
\end{equation} 
(see, e.g., Shapiro \& Teukolsky 1983, sec 12.4). 
Here, $\ell (\equiv r v_\varphi)$ denotes the specific angular momentum 
of the test particle and the last term on the right-hand side
corresponds to 
the potential of the centrifugal force. 
The general relativistic effect appears in the term in proportion to $r^{-3}$.
It is this term that makes mass accretion to a black hole possible,
even when matter has a finite angular momentum.
A particle having angular momentum ($\ell$) greater than the critical
value, $\ell_{\rm crit}$ (which is on the order of $c r_{\rm g}$), 
can rotate on a stable circular orbit  at a certain radius determined by 
 the minimum of the effective potential of constant $\ell$. 
As $\ell$ decreases, this equilibrium radius decreases and,
when $\ell < \ell_{\rm crit}$, it disappears.  
Thus, the minimum value of the equilibrium radius 
is $3~r_{\rm g}$ for a non-rotating black hole. 
This argument is directly applicable to the case of general
relativistic fluid (Lu et al. 1995),
 and the marginally stable orbit in such a case is, again,
 $\approx 3~r_{\rm g}$. 

In short, the stability argument assumes the presence of 
 an equilibrium point, in which all the forces are balanced, 
 for a given angular momentum.  A problem arises, when the inertial
 term [$\propto v_r (dv_r/dr)$] becomes substantial.  
This indeed occurs around the transonic point,
 in which matter velocity equals to sound velocity. 
Since the flow speed is $\sim c$ while the sound velocity is at most 
$\sim c/\sqrt{3}$ near the event horizon, the flow should be
 supersonic at small radii but must be subsonic at large radii;
 that is, there should be a transonic point near the black hole
 (see Chap. 9 of Kato, Fukue, \& Mineshige 1998 for more details).

Again, the problem is not serious in standard-type disks with 
luminosity, $L < L_{\rm E}$,
in which the inner edge has been shown to be barely inside the
marginally stable orbit at $r_{\rm ms}$
(Muchotrzeb \& Paczy{\'n}sky 1982; Muchotrzeb 1983; Matsumoto
et al. 1984; Abramowicz \& Kato 1989).  
A noteworthy effect appears in
optically thin ADAF (advection-dominated accretion flow) and in
near-critical accretion flow (noted as a slim disk).
Although transonic nature of the flow has been extensively discussed mainly
in the former case (Narayan, Honma, \& Kato 1997)
and in the context of geometrically thick torus,
the place of the inner edge of the flow has been poorly investigated
so far in relation to the stability of circular orbits.


\subsection{Basic equations and numerical procedures}
The numerical procedures are the same as those adopted by Watarai et al. 
(2000, see Chap. 10 of Kato, Fukue, \& Mineshige 1998 
 for the derivations of the basic equations and the detailed discussion). 
The essential ingredients are the transonic condition of the flow 
 and the presence of advective energy transport.  
In the present study we adopt the pseudo-Newtonian potential 
(Paczy{\'n}sky \& Wiita 1980) to simplify the treatment.  
It is known that this potential is a good approximation of the general
relativistic one down to $\sim 2~r_{\rm g}$ and gives the correct
radius of the marginally stable orbit (see section 2.2).  
Thus, the adoption of the pseudo-Newtonian potential will not
introduce serious errors when we are concerned with the properties of
the flow at $r \gtrsim 3~r_{\rm g}$. 

The basic equations are as follows:
The mass conservation reads
\begin{equation}
\label{mass}
  {\dot M} = -2\pi r\Sigma v_r,
\end{equation}
where $v_{r}$ and $\Sigma$ are radial velocity and surface density, 
respectively.
The momentum equation in the radial direction is  
\begin{equation}
\label{r-mom}
   v_r{dv_r\over dr}+{1 \over \Sigma}{dW\over dr}
   ={\ell^2-\ell_{\rm K}^2\over r^3}
   -{W\over \Sigma}{d{\rm ln}~\Omega_{\rm K}\over dr},   
\end{equation}
where $W \equiv \int p_{\rm tot} dz$ is the integrated pressure and
the Keplerian angular momentum is $\ell_{\rm K}=r^2\Omega_{\rm K}$,
 with $\Omega_{\rm K}$ being the Keplerian angular frequency under
 the pseudo-Newtonian potential. 
Note that the last term on the right-hand side is a correction term
for gravity. 

The angular momentum conservation is written as
\begin{equation}
\label{ang-mom}
    {\dot M}(\ell-\ell_{\rm in})=-2\pi r^2T_{r\varphi},  
\end{equation}
where
$T_{r\varphi}$ is the height integrated viscosity stress tensor
and we adopt the $\alpha$ viscosity prescription;
\begin{equation}
\label{viscosity}
  T_{r\varphi} \equiv -\alpha p_{\rm tot} H.
\end{equation}
Here, $H$ denotes the half-thickness of the disk and is
\begin{equation}
\label{h-balance}
    H = (p_{\rm tot}/\rho \Omega)^{1/2}
\end{equation}
 from the hydrostatic balances in the vertical direction.
The equation of state is 
\begin{equation}
\label{pressure}
   p_{\rm tot}= p_{\rm gas} + p_{\rm rad} = 
   \frac{\Re}{\bar{\mu}} \rho T + \frac{a}{3}T^4,    
\end{equation}
where $\Re, \bar{\mu}$ and $a$ are the gas constant, 
mean molecular weight, and the radiation constant, respectively.
Note that all the quantities refer to the values
on the equatorial plane. 

The energy equation involves viscous heating, 
radiative cooling, and advective cooling;
\begin{equation}
\label{energy}
  Q_{\rm vis}^{+}=Q_{\rm rad}^{-}+Q_{\rm adv}^{-}
\end{equation}
where each term is explicitly written as
\begin{eqnarray}
\label{q_vis}
    {Q_{\rm vis}^+} &=& r T_{r\varphi}\frac{d\Omega}{dr} \\  
\label{q_rad}
    {Q_{\rm rad}^-} &=& \frac{8 a c T^4}{3 \bar{\kappa} \rho H}, \\  
\label{q_adv}
    {Q_{\rm adv}^-} &=& \frac{9}{8} v_r\Sigma T \frac{ds}{dr}. 
\end{eqnarray}
Here, $s$ denotes specific entropy in the equatorial plane and
$\bar{\kappa}$ is the Rosseland-mean opacity.

We solved the set of equations (\ref{mass})--(\ref{energy}) 
from the outer boundary at $10^4~r_{\rm g}$ inward down to $\sim r_{\rm g}$
through the transonic point.  The inner boundary is free.

\subsection{Effective potentials of high $\alpha$ disks}
The expression for the effective potential is then
\begin{equation}
\psi_{\rm eff}^{\rm PN0}(r) = -\frac{GM}{r-r_{\rm g}}+\frac{\ell^2}{2r^2}.
\label{eq:p1}
\end{equation}
If we fixed $\ell$, the condition of 
 $d\psi_{\rm eff}(r)/dr=0$ usually gives three solutions, among which
only one solution is stable, since it corresponds to a potential minimum,
$d^2\psi_{\rm eff}(r)/dr^2>0$, and the other two are unstable solutions
with $d^2\psi_{\rm eff}(r)/dr^2<0$.
We vary $\ell$ and seek for a solution which satisfies
the condition, $d^2\psi_{\rm eff}(r)/dr^2=0$, at the radius where
$d\psi_{\rm eff}(r)/dr=0$ holds.
We find the solution for $\ell_{\rm crit}=1.837$ and
the derived radius of the marginal stable orbit is $r_{\rm ms}=3~r_{\rm g}$, 
just the same as that derived based on the Schwarzschild metric. 
When the angular momentum of a test particle is less than this critical value, 
there no longer exists a stable circular orbit.

 \begin{figure}[htb]
  \begin{center}
    \FigureFile(80mm,80mm){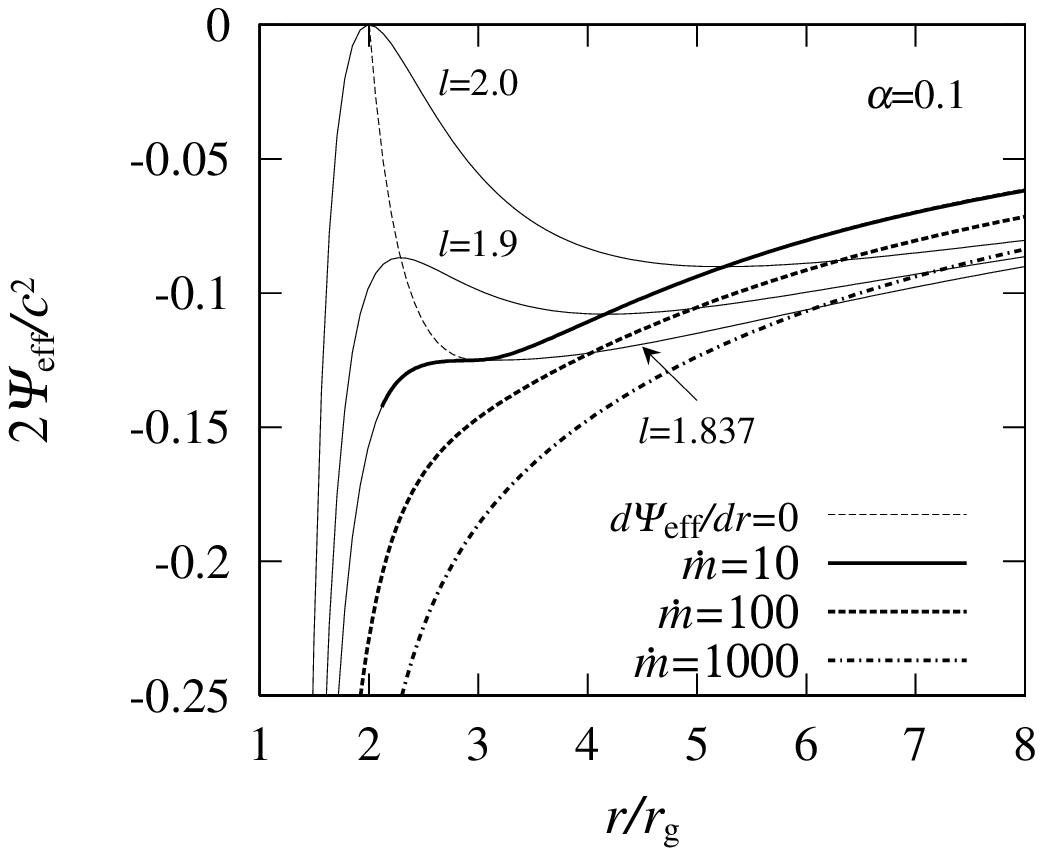}
    \FigureFile(80mm,80mm){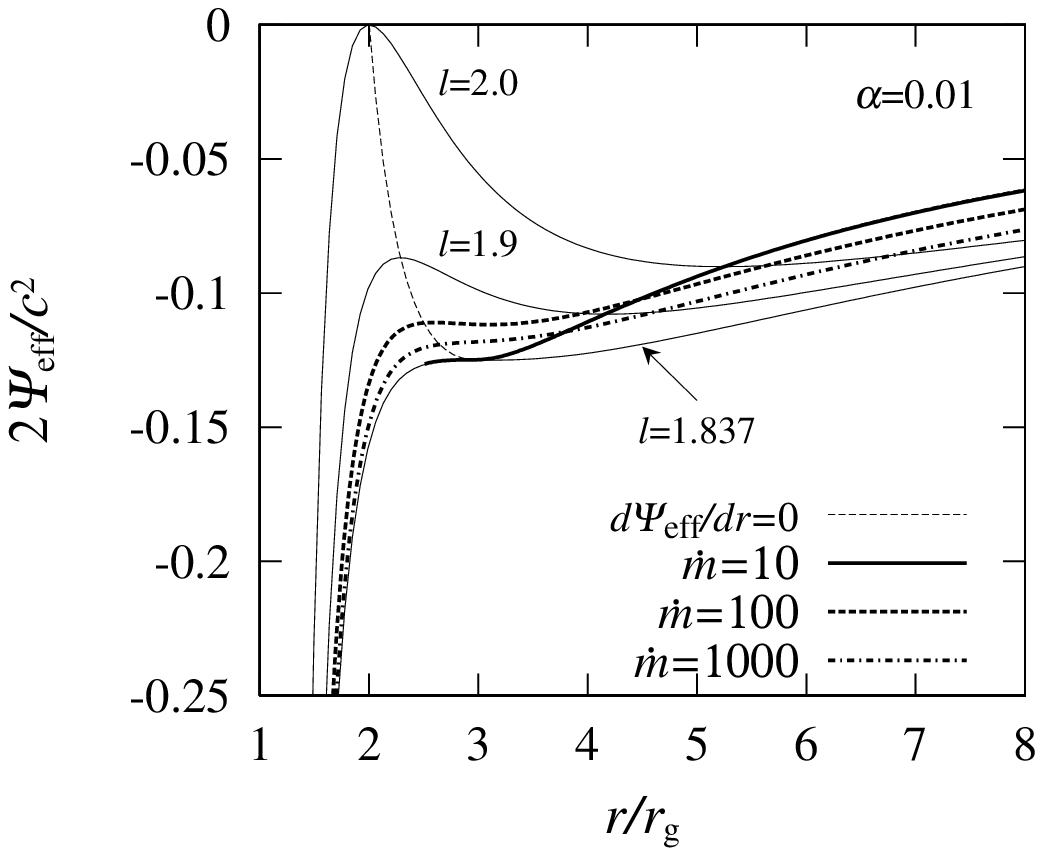}
  \end{center}
  \caption{Profiles of the effective potentials for $\alpha=0.1$ 
(upper panel) and $\alpha=0.01$ (lower panel), respectively.  
The thick solid, dashed, dotted lines represent the numerically
   calculated potential for accretion rates of $\dot{m}
 (\equiv \dot{M}c^2/L_{\rm E})=$ 10, 100, and 1000 from the top to bottom.
 The thin solid lines show the same but for a fixed specific angular momentum
 of $\ell (\equiv r v_{\varphi})$ = 2.0, 1.9, and
 1.837 from the top to bottom. 
 The thin dotted line represents $\psi_{\rm eff}=0$ curve.
}
  \label{fig:effpot1}
\end{figure}

Next, we consider the case of fluid
and calculate the effective potential based on the numerical models. 
We adopt the numerical value of specific angular momentum,
which is a function of radius, to calculate the effective potential.
 Figure \ref{fig:effpot1} shows the radial profile of the
effective potential, $\psi_{\rm eff}^{\rm PN0}$, for various accretion rates,
$\dot{m}=10, 100$, and $1000$.  Here, we define the normalized 
accretion rate to be $\dot{m} \equiv \dot{M}c^2/L_{\rm E}$.
We fix the black hole mass to be $M =10~M_{\odot}$
in the entire calculations. 
For comparison, we also plot the effective potentials of fixed
angular momenta, $\ell$ = 2.0, 1.9 and 1.837 ($=\ell_{\rm crit}$).
The thin dotted line represents the locus of the stable orbits;
namely, we connect the points
where $d\psi_{\rm eff}/dr = 0$ and $d^2\psi_{\rm eff}(r)/dr^2>0$
hold for various values of $\ell$ ($\geq \ell_{\rm crit}$).
This is the trajectory along which accreting gas would move
if the gas had the Keplerian rotation velocity
(i.e., if the inertial and pressure terms were zero; see
equation [\ref{r-mom}]).

Let us compare the numerical solutions with the analytical ones (fixed
$\ell$ solutions)  
for the cases with relatively high $\alpha=0.1$ (see the upper panel).
We notice that both well agree for small accretion rates with
$\dot{m}=$1 and 10.  This result justifies the usual belief that
the inner edge of the disk is determined by the marginal stable orbit.
The sonic points lie slightly inside $3~r_{\rm g}$,
as was already pointed out (Matsumoto et al. 1984).  
For large accretion rates, $\dot{m}\geq 100$, in contrast,
big discrepancies appear.
The rotation velocity is significantly lower than the Keplerian value,
which causes the downward shift of the potential profile.
The larger $\dot{m}$ is, the lower becomes the value of the potential,
indicating a higher efficiency of angular-momentum removal.
Due to this reason, such an accretion flow is called as viscosity-driven flow.
As a consequence, we see $\ell < \ell_{\rm crit}$ at
$r \lesssim 4~r_{\rm g}$ for ${\dot m} = 10$
and at $r \lesssim 6~r_{\rm g}$ for ${\dot m} = 100$;
i.e., always gravity dominates over the centrifugal force
and the inertial term gives an important contribution
in the momentum equation.
Since there is no minimum in their effective potential profile, 
there is no equilibrium place and, hence,
the usual technique of the stability analysis cannot be applied here.
Note that our derived angular-momentum distribution agrees well
with those found by
Abramowicz et al. (1988) and Artemova et al. (2001), but
they did not show the effective potential profiles.

\subsection{Effective potentials of low $\alpha$ disks}
Let us turn to the cases of small $\alpha = 0.01$ (see the lower panel
of figure 1).
The behavior of the low $\dot m$ disk does not differ from the previous cases,
but that of high $\dot m$ disk is qualitatively different;
most notably, the angular momentum at $3~r_{\rm g}$ still exceeds
the critical value.   
This means that the rotation velocity is super-Keplerian.
Nevertheless, accretion is possible because of 
enhanced pressure gradient force.
In fact, we find a pressure peak at $4 \sim 5~r_{\rm g}$ and, hence,
the pressure gradient force can induce gas accretion flow 
at $r \lesssim 4~r_{\rm g}$.  
Hence, this kind of flow is called as pressure-driven flow.
The potential profile has a minimum, even when $\dot{m} \geq 100$,
which makes a good contrast with the low $\alpha$ disks.  
The radius of the potential minimum is around $3~r_{\rm g}$.
However, we cannot simply conclude that an orbit at this radius
is stable, since matter cannot stay at this radius.
In other words, this is not an equilibrium radius.

\subsection{Effect of pressure gradient}

We have basically confirmed results of previous authors at low
accretion rates \footnote{We also calculated $\dot{m}$=1, 0.1, 0.01 
cases, however, the shape of the effective potential is almost same as
$\dot{m}=10$.}. 
In addition, we find that accretion rates are another key parameter 
to determine the types of the accretion processes.  
Note, however, that equation (\ref{eq:p1}) does not include the effect
of pressure gradient. 
Hence, we modify the effective potential as,   
\begin{equation}
\psi_{\rm eff}^{\rm PN1}(r) = -\frac{GM}{r-r_{\rm
g}}+\frac{\ell^2}{2r^2}-\int^{r_{\rm out}}_{r}
\frac{1}{\Sigma}\left(\frac{dW}{dr}\right) dr.
\label{eq:p2}
\end{equation}
Here, the third term describe the change of the internal energy by
word done by the pressure gradient force 
\footnote{We ignored the correction for gravity, 
since its contribution is small.}.
This term is important to understand the behavior of a fluid element 
 because the flow becomes radiation pressure dominated under high mass
accretion rate. 

\begin{figure}[htb]
  \begin{center}
   \FigureFile(80mm,80mm){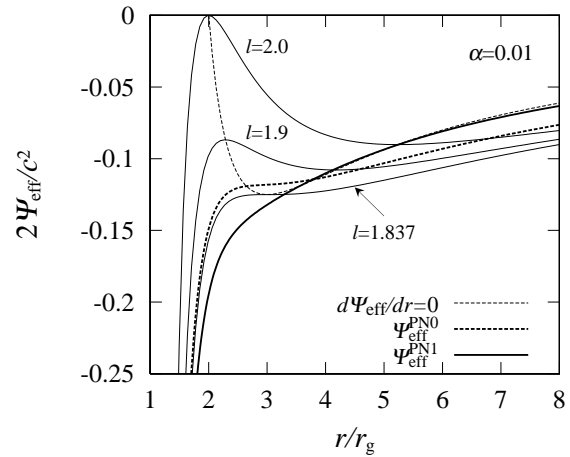}
  \end{center}
  \caption{Profiles of the effective potentials for $\alpha=0.01$.  
The thick solid lines represent the numerically calculated potential,
   equation (\ref{eq:p2}),
for accretion rates of $\dot{m} (\equiv \dot{M}c^2/L_{\rm E})=$ 1000.
 The thick dotted lines show the calculated potential, but the lines
 are derived from equation (\ref{eq:p1}). 
}
  \label{fig:effpot2}
\end{figure}

Figure $\ref{fig:effpot2}$ shows the contributions of pressure
gradient term for $\dot{m}=1000$ case. 
The thick dotted lines on figure \ref{fig:effpot2} are calculated by
equation (\ref{eq:p2}); i.e., the effect of pressure gradient is excluded. 
In such cases, we can distinguish the flow type with the $\alpha$ parameters. 
For large $\alpha$ (upper panel in figure $\ref{fig:effpot2}$), 
 the flow is driven by viscosity then there is no potential minimum. 
On the other hand, for smaller $\alpha$
 (lower panel in figure \ref{fig:effpot2}),
 the accretion flow inside $\sim$ 3 $r_{\rm g}$ is driven by pressure so
that there is a potential minimum even if relatively high mass accretion rate. 
In both cases, however, if we include the pressure term (i.e.,
the thick solid lines, which are derived from equation (\ref{eq:p2})),
 then the effective potential does not have minimum. 
Pressure gradient term is not so important for large $\alpha$ case, 
because the flow is basically viscosity-driven. 
However, for small $\alpha$ case, the form of the effective potential
significantly changed due to the effect of pressure gradient. 
To induce accretion onto black hole for small $\alpha$, 
 accretion must be driven by the pressure gradient force, and the
disk must have a pressure peak in the inner region. 
Inside of the pressure peak, the pressure gradient force works inward. 
Conversely, the pressure gradient force works outward outside
 the pressure peak. 
Hence, the pressure minimum does not appear even for small $\alpha$
model \footnote{We also confirm that there is no radius where
$\psi_{\rm eff}/dr=0$ holds, for smaller $\alpha$(=0.001) case. }. 
Again, we note that the classical analysis based on equation
(\ref{eq:p0}) or (\ref{eq:p1}) is 
no longer valid in high mass-accretion systems 
because of the significant effect of radiation pressure force.

\subsection{Viscosity driven flow and pressure driven flow}
 Transonic nature of optically thick accretion disk has been extensively
investigated since the early 1980's. 
Abramowicz et al. (1978) found the cusp structure located between $r_{\rm ms}$
and marginally bound orbit, $r_{\rm mb}$,
 for any angular momentum distribution. 
They proposed that mass accretion into a black hole may occur in a 
similar way to the case of the Roche-lobe overflows.
Muchotrzeb \& Paczy{\'n}sky (1982) were the first to 
find numerically that the cusp radius is
located between $r_{\rm ms}$ and $r_{\rm mb} (\sim 2~r_{\rm g})$, 
for a low viscosity parameter, $\alpha=0.001$. 
Their conclusion is that the ``no torque boundary
condition at $r_{\rm in}$'' is valid. 
 Matsumoto et al. (1984) have found distinct flow structure
in cases with high viscosity parameters and claimed that
flow properties differ, depending on the magnitudes of viscosity.
For small $\alpha$ ($\lesssim 0.05$) the pressure peak around 
$r \simeq 4-5~r_{\rm g}$ drives the mass accretion, whereas for
large $\alpha$ ($\gtrsim 0.05$) the flow is mainly driven by viscosity.  

\begin{figure}[htb]
  \begin{center}
   \FigureFile(80mm,80mm){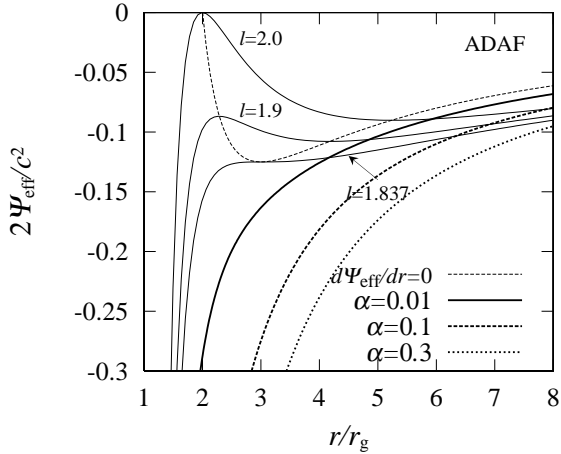}
  \end{center}
  \caption{Same as figure 1 but for the optically thin ADAF.
The solid lines represent the effective potentials for
  $\alpha$= 0.01, 0.1, 0.3 from the top to bottom.  
We set the accretion rate to be $\dot{m}=10^{-3}$. 
There is little accretion rate dependence.
}
  \label{fig:adaf}
\end{figure}

We also calculate optically thin ADAF with different $\alpha$ parameters
and display the resultant effective potential in figure \ref{fig:adaf}. 
In ADAF, we simply assume one-temperature plasma and bremsstrahlung
emission as the cooling process. 
Other equations are the same as those of the optically thick case. 
Unlike the case of optically thick disk,
the effective potential profile of the optically thin ADAF does not
exhibit strong $\dot m$-dependence.
Our result is in good agreement with those by Narayan et al. (1997). 

\section{Discussion}
We have demonstrated that 
a marginally stable orbit does not always exist in high luminosity disks;
that is, the inner disk radius is not necessary equal to $3~r_{\rm g}$.
 This indicates that the inner disk radius derived by the 
fitting to the observed soft X-ray spectra 
based on the multi-color disk model (Mitsuda et al. 1984)
does not always correspond to $r_{\rm ms}$, especially at high $\dot m$.
The same is true for the ADAF. 
There should be no problems, however, for low luminosity,
standard-type disks with $L \lesssim L_{\rm E}$. 

Our results should influence the observational interpretation as to the
black hole rotation (Zhang et al. 1997; Makishima et al. 2000). 
The widely used argument is that if the inner-disk radius 
estimated by spectral fitting etc. is significantly less than $3~r_{\rm g}$,
the black should be a Kerr hole.  
This argument does not hold for luminous systems shining at $L \sim
L_{\rm E}$,
 since they always show a smaller inner-edge radius (Watarai et al. 2000,2001).

The same is true for the optically thin ADAF.
The asymmetric iron K$\alpha$ line profiles observed at 6.4 keV in black
hole candidates provides a plausible evidence of strong
gravitational field of black hole (Fabian et al. 1989; Kojima 1991;
Tanaka et al. 1995; Miller et al. 2002),
and from the spectral fitting
it has been claimed that the central black hole should be rapidly rotating
(Iwasawa et al. 1999), since
the inner disk radius derived by the fitting is less than $3~r_{\rm g}$. 
We, however, argue that
the inner edge in optically thick accretion disk,
$r_{\rm in}$, can be smaller than $3~r_{\rm g}$ in the cases with
large $\alpha$ and high $\dot M$.  
Since the effect of gravitational redshift is on the
 order of $1/\sqrt{1-r_{\rm g}/r}$, 
it might not be so easy to discriminate the effects of
black hole rotation and those of super-critical flow
solely from the spectral lines and/or the thermal spectral component.
 
 Recently, several groups have succeeded in performing multi-dimensional
magneto-hydrodynamic (MHD) simulations of accretion flow and
confirmed the presence of a marginally stable orbit 
around 3$r_{\rm g}$ (Hawley \& Balbus 1999; Hawley 2000;
Armitage et al. 2001; Hawley \& Balbus 2002). 
On the other hand, Krolik \& Hawley (2002) proposed the ``inner edge''
of an accretion disk around a black hole depends on the definition,
i.e., turbulence edge, stress edge, reflection edge and radiation edge
are defined by each physical process. 
That is, the often used expression ``inner edge'' does not always
correspond to $r_{\rm ms}$.  
This paper describes one such example.

\section{Conclusions}
 In this paper, we discuss the dynamical properties of the
transonic flow in super-critical regimes ($\dot{M} \gtrsim 10 L_{\rm E}/c^2$)
and demonstrate that the marginally stable orbit does not exit
for high mass accretion rates ($\dot{m} \gtrsim 10$). 
For small viscosity parameters ($\alpha \lesssim 0.01$), 
the effect of pressure gradient is important for discussion of
 the marginal stable orbit, 
since the effective potential including the pressure gradient
 does not have a minimum. 
That is, the usual argument based on the assumption
that the inner edge of the disk coincides with the 
marginally stable orbit does not work here.
The inner-disk radius obtained by the spectral fitting does not
necessarily coincide with the radius of the marginally stable orbit.
The observed value is more like the so-called radiation edge,
outside which substantial emission is produced.  We cannot simply
relate the marginal stable orbit with the inner boundary.
To determine the radiation edge precisely, we need to solve full radiation
transfer equations in more than two dimensions, which is left as future work.

\vspace{10mm}
We would like to thank the referee M.A. Abramowicz for his critical comments
and suggestions. 
We are also grateful to S. Kato, W. Kluzniak, J. Fukue, K. Ohsuga, Y. Kato, 
and R. Takahashi for their useful comments and discussions.   
This work was supported in part by the Grants-in Aid of the
Ministry of Education, Science, Sports, and Culture of Japan
(14001680, KW; 13640328,14079205, SM).

 
\end{document}